\documentclass[11pt,preprint]{aastex}
\newcommand{\teff}{\mbox{$T_{\rm eff}$}}
\newcommand{\logg}{\mbox{$\log g$}}
\newcommand{\vsini}{\mbox{$v \sin i$}}
\newcommand{\mictrb}{\mbox{$\xi_{\rm t}$}}

\newcommand{\kms}{\mbox{km\, s$^{-1}$}}

\shorttitle{WASP-19b: the shortest period exoplanet yet discovered}
\shortauthors{Hebb et al.}

\begin{document}

%%%%%%%%%%%%%%%%%%%%%%%%%%%%%%%%%%%%%%%%%%%%%%%%

\title{WASP-19b: the shortest period transiting exoplanet yet discovered}

\author{L.~Hebb\altaffilmark{1},
A.~Collier-Cameron\altaffilmark{1},
A.H.M.J.~Triaud\altaffilmark{2},
T.A.~Lister\altaffilmark{3},
B.~Smalley\altaffilmark{4},
P.F.L.~Maxted\altaffilmark{4},
C.~Hellier\altaffilmark{4},
D.R.~Anderson\altaffilmark{4},
D.~Pollacco\altaffilmark{5},
M.Gillon\altaffilmark{2,6},
D.~Queloz\altaffilmark{2},
R.G.~West\altaffilmark{7},
S.Bentley\altaffilmark{4},
B.~Enoch\altaffilmark{1},
C.A.~Haswell\altaffilmark{8},
K.~Horne\altaffilmark{1},
M.~Mayor\altaffilmark{2},
F.~Pepe\altaffilmark{2},
D.~Segransan\altaffilmark{2},
I.~Skillen\altaffilmark{9},
S.~Udry\altaffilmark{2},
\and
P.J.~Wheatley\altaffilmark{10}
}

\altaffiltext{1}{School of Physics and Astronomy, University of St Andrews, North Haugh, St Andrews, Fife KY16 9SS, UK }
\altaffiltext{2}{Observatoire de Gen\`eve, Universit\'e de Gen\`eve, 51 Ch. des Maillettes, 1290 Sauverny, Switzerland }
\altaffiltext{3}{Las Cumbres Observatory, 6740 Cortona Dr. Suite 102, Santa Barbara, CA 93117, USA }
\altaffiltext{4}{Astrophysics Group, Keele University, Staffordshire, ST5 5BG, UK }
\altaffiltext{5}{Astrophysics Research Centre, School of Mathematics \&\ Physics, Queen's University, University Road, Belfast, BT7 1NN, UK}
\altaffiltext{6}{Institut d'Astrophysique et de Géophysique, Université de Liège, Allée du 6 Août, 17, Bat. B5C, Liège 1, Belgium}
\altaffiltext{7}{Department of Physics and Astronomy, University of Leicester, Leicester, LE1 7RH, UK }
\altaffiltext{8}{Department of Physics and Astronomy, The Open University, Milton Keynes, MK7 6AA, UK}
\altaffiltext{9}{Isaac Newton Group of Telescopes, Apartado de Correos 321, E-38700 Santa Cruz de la Palma, Tenerife, Spain }
\altaffiltext{10}{Department of Physics, University of Warwick, Coventry CV4 7AL, UK}
%\altaffiltext{2}{Laboratoire d'Astrophysique de Marseille, BP 8, 13376 Marseille Cedex 12, France }
%\altaffiltext{4}{Institut d'Astrophysique de Paris, CNRS (UMR 7095) --  Universit\'e Pierre \&\ Marie Curie, 98$^{bis}$ bvd. Arago, 75014 Paris, France }
%\altaffiltext{6}{Observatoire de Haute-Provence, 04870 St Michel l'Observatoire, France }
%\altaffiltext{10}{School of Physics, University of Exeter, EX4 4QL, UK }
%\altaffiltext{12}{Harvard-Smithsonian Center for Astrophysics, 60 Garden Street, Cambridge, MA, 02138 USA}

\begin{abstract}
We report on the discovery of a new extremely short period transiting extra-solar planet, WASP-19b.
The planet has mass, $M_{\rm pl}=1.15 \pm 0.08$~M$_J$, radius, $R_{\rm pl}=1.31 \pm 0.06$~R$_J$, 
and orbital period, $P=0.7888399 \pm 0.0000008$~days.  Through spectroscopic analysis,
we determine the host star to be a slightly super-solar metallicity ($[M/H]=0.1 \pm 0.1$~dex) G-dwarf 
with $T_{\rm eff}=5500 \pm 100$~K.  In addition, we detect periodic, sinusoidal flux variations 
in the light curve which are used to derive a rotation period for the star
of $P_{rot}=10.5 \pm 0.2$~days.  The relatively short stellar rotation period suggests
that either WASP-19 is somewhat young ($\sim 600$~Myr old) or tidal interactions between the two bodies
have caused the planet to spiral inward over its lifetime resulting in the spin-up of the star.
Due to the detection of the rotation period, this system has the potential to place strong
constraints on the stellar tidal quality factor, $Q_{s}^{\prime}$, if a more precise age is determined.
\end{abstract}


\maketitle

%______________________________________

\section{Introduction}

Since the unexpected discovery of the first `hot Jupiter', 51~Peg~b \citep{peg}, exoplanets
with an exceptionally wide variety of properties have been detected which have dramatically changed
our understanding of planetary physics.  In particular, through the discovery of various transiting planets,
we have learned that extra-solar planets can have radii much larger than Jupiter \citep[e.g.][]{wasp12} or densities much
higher \citep{hotsat}.  Many, but not all, `hot Jupiters' have temperature inversions
in their atmospheres \citep[e.g.][]{tempinvert}, and they can have very low optical albedos \citep{Rowe07}.  Despite their short periods, not all transiting exoplanets have been tidally 
circularized \citep{wasp6}, and both
rocky (e.g. CoRoT-Exo-7, $P\sim0.85$~days) and gas giant (e.g. WASP-12b, $P\sim1.09$~days) 
planets can exist in extremely short period orbits.
Here, we report on the discovery of a new extreme transiting extra-solar planet with the shortest
orbital period yet detected which is on the verge of spiraling into its host star.  
This transiting planet can not only inform us about the properties and evolution of close-in planets, 
but it also has the potential to provide information about the characteristics of its host star.

In this paper, we first describe all the observations that were obtained to
detect and analyse the transiting star-planet system (\S\ref{sec:observations}).
We describe the data analysis in \S\ref{sec:analysis}
where we present the planet and its host star.
Finally in \S\ref{sec:discuss}, we discuss the
implications of the planet's short period and its future evolution.

\section{Observations}
\label{sec:observations}

2MASS~J09534008-4539330 (hereafter WASP-19) is an apparently unremarkable 12th magnitude ($V=12.59$), G8V star 
in the southern hemisphere located at $\alpha=$09:53:40.08, $\delta=$-45:39:33.0 (J2000).  
The target was observed with the WASP-South telescope and instrumentation  \citep{swasp_instr,wasp4}
in the winter and spring observing seasons from 2006 to 2008.  1496 photometric
data points were obtained between 4 May - 20 June 2006, 6695
measurements were made between 18 Dec 2006 - 18 May 2007, 
and 8968 observations were taken from 18 Dec 2007 - 22 May 2008.  
All data sets were processed independently with the standard WASP data reduction
pipeline and photometry package \citep{hunter}.  The individual data points
have typical uncertainties of $\sim 0.02$~mags including poisson noise
and systematic noise.   The resulting light curves were then
run through our implementation of the box least squares algorithm \citep{kovacs} 
designed to detect periodic transit-shaped dips in brightness.

\begin{figure}
  \centering
  \includegraphics[angle=0,width=\columnwidth]{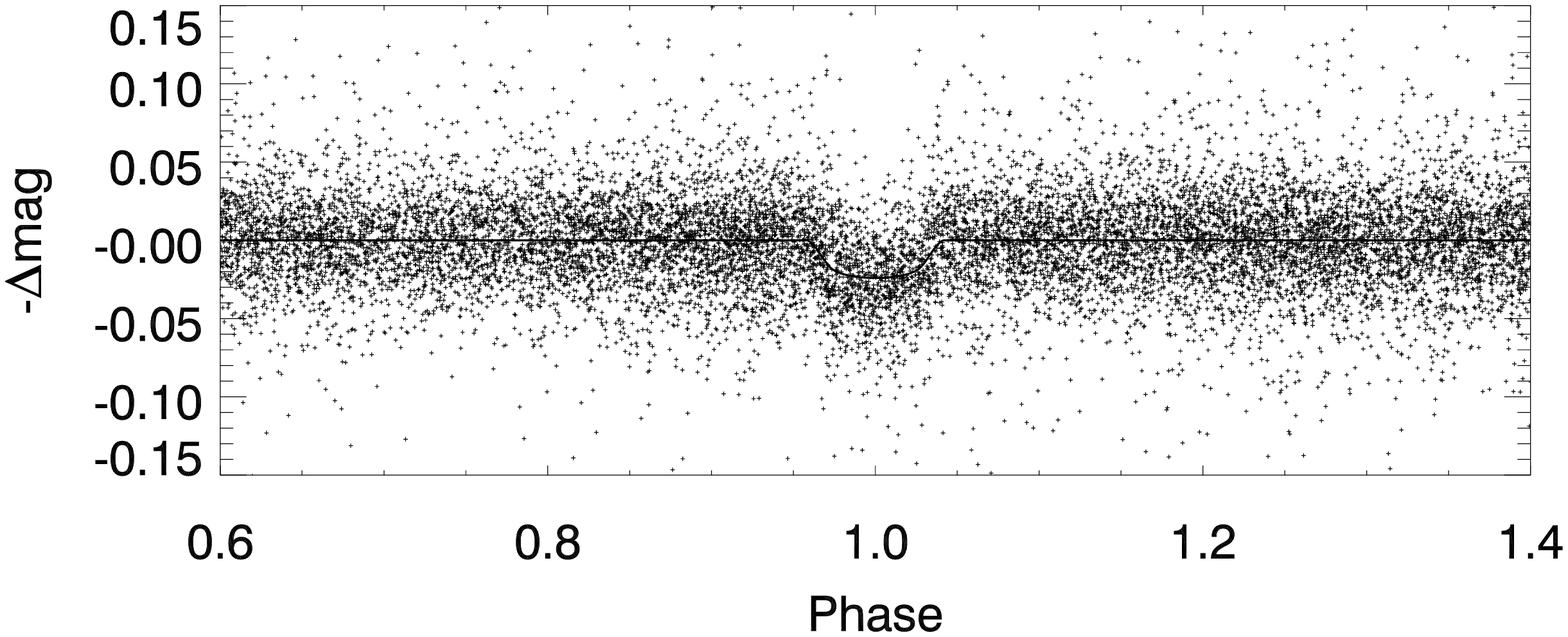}
  \caption{WASP-South discovery photometry of WASP-19.  The data 
   are phase-folded with the ephemeris given in Table~\ref{tab:params}.
%   Overplotted are the best fitting model transit light curves using the formalism
%   of \citet{mandelagol} applying the 4th-order limb darkening coefficients from \citet{claret00}.
  }
  \label{fig:swaspphot}
\end{figure}

The target was initially flagged as a transiting planet candidate because a strong
periodic signal was detected in the the 2007 data.  The phase-folded light curve
showed a square-shaped dip in brightness with a depth, $\delta \sim 25$~mmag
and duration, $\tau \sim 1.2$~hours, consistent with a planet sized object around
a main sequence star.  Further, a periodic transit was also apparent in the 2006 data when 
phase-folded with the 2007 ephemeris, and a transit was subsequently detected in the 2008 season of data.  
Therefore, we classified the object as needing follow up photometry
and spectroscopy to assess the planetary nature of the system.
The phase-folded light curve containing all WASP-South data is shown 
in Figure~\ref{fig:swaspphot}.  

\begin{figure}
  \centering
  \includegraphics[angle=0,width=\columnwidth]{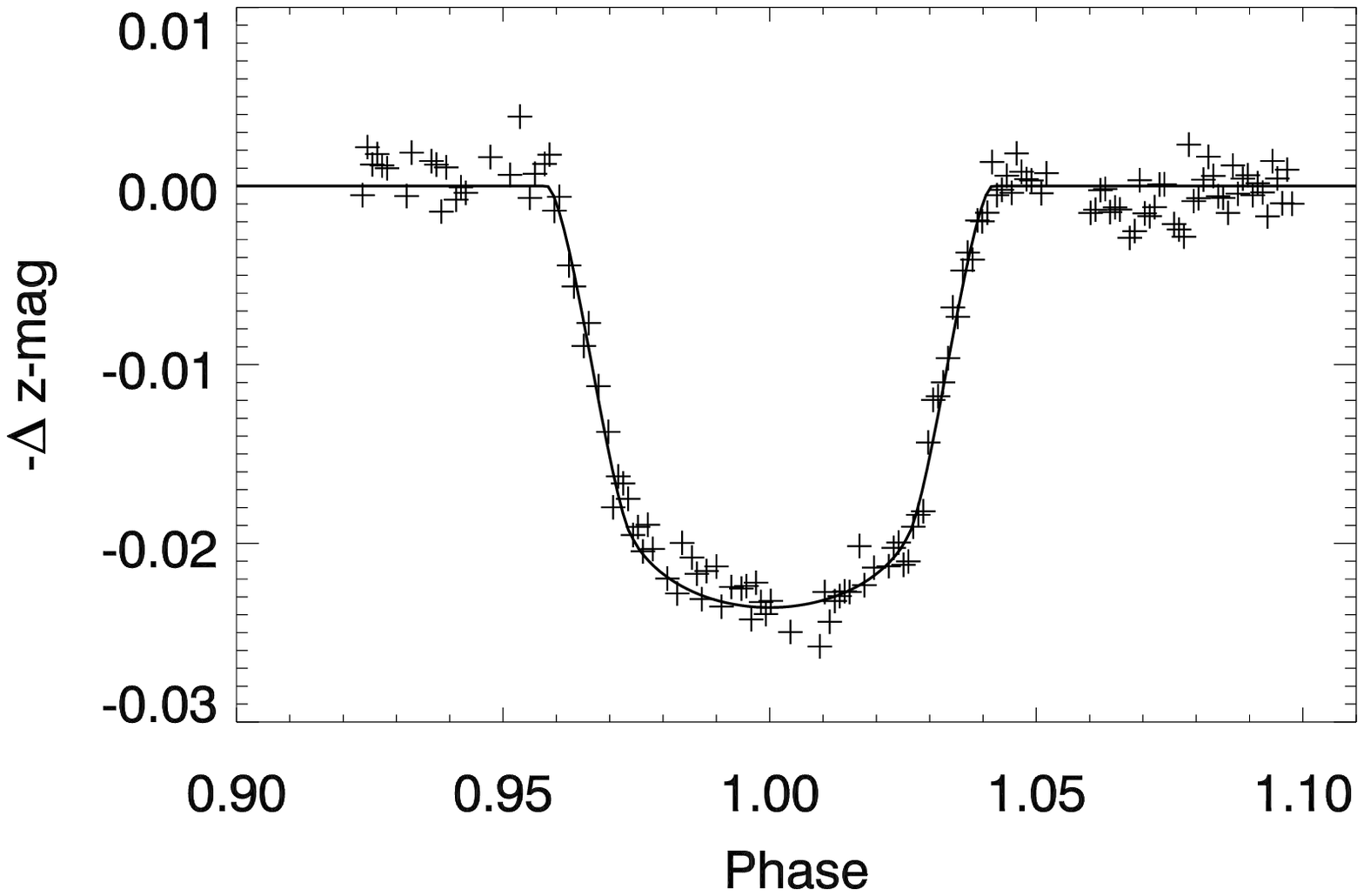}
  \caption{FTS $z$-band photometry of the WASP-19 transit.  The data are converted
   to phase using the ephemeris given in Table~\ref{tab:params}.
   Overplotted is the best fitting model transit light curve using the formalism
   of \citet{mandelagol} applying the 4th-order limb darkening coefficients from \citet{claret04}.
  }
  \label{fig:phot}
\end{figure}

WASP-19 was observed photometrically with the 2m Faulkes Telescope South (FTS) 
on 17 December 2008 during
transit.  139 Pan-STARRS z-band\footnote{http://pan-starrs.ifa.hawaii.edu/public/design-features/cameras.html} 
observations were made over 3.3~hours.  The images were observed
in $2\times2$~binning mode such that one binned pixel corresponds to $0.279^{\prime\prime}$.  
They were processed in the standard way with IRAF using a stacked bias image, dark frame, and sky flat.  
Minimal fringing was present in the z-band images due to the deep depletion CCD in the camera, so
no fringe correction was applied.
The DAOPHOT photometry package \citep{daophot} was used to perform object detection and
aperture photometry with an aperture size of 8~binned-pixels in radius.
The $5^{\prime}\times 5^{\prime}$ field-of-view of the instrument contained 53 comparison stars 
that were used in deriving the differential magnitudes with a photometric
precision of 1.3~mmag.  We measured the red noise \citep{Pont06} in the light curve on a 
30~minute timescale to be 449 ppm and added this value in quadrature to the formal uncertainties on each
data point.  The resulting light curve is shown in Figure~\ref{fig:phot}.  

\begin{table}
\caption[]{Radial velocity measurements of WASP-19 obtained with the CORALIE spectrograph.  }
\label{tab:rvdata}
\begin{center}
\begin{tabular}{ccccc}
BJD   & $\rm{V}{r}$  & $\rm{\sigma_{RV}}$  & Bisector \\
     &  km s$^{-1}$ & km s$^{-1}$          & km s$^{-1}$\\
\hline\\
2454616.46326  & 20.965  &   0.0218 & -0.02975  \\
2454623.46713  & 20.712  &   0.0289 & 0.03735    \\
2454624.46191  & 21.019  &   0.0220 & 0.00607    \\
2454652.46587  & 20.512  &   0.0210 & 0.02176  \\
2454653.46543  & 20.770  &   0.0231 & -0.03106  \\
2454654.46530  & 21.007  &   0.0219 & 0.03017    \\
2454656.47564  & 20.511  &   0.0193 & -0.04093  \\
2454657.46805  & 20.902  &   0.0386 & 0.04432    \\
2454658.46376  & 21.008  &   0.0302 & 0.05792    \\
2454660.46652  & 20.604  &   0.0400 & 0.05038    \\
2454661.46434  & 20.974  &   0.0217 & 0.04052    \\
2454662.46549  & 20.922  &   0.0191 & 0.01634    \\
2454663.46607  & 20.547  &   0.0188 & -0.02782  \\
2454664.46573  & 20.645  &   0.0362 & -0.14578  \\
2454665.46709  & 21.077  &   0.0305 & -0.00590  \\
2454827.74752  & 20.683  &   0.0188 & 0.02320    \\
2454832.74362  & 21.050  &   0.0242 & -0.08001  \\
2454833.66844  & 20.860  &   0.0212 & 0.01644    \\
2454834.67693  & 20.548  &   0.0191 & -0.04711  \\
2454837.67024  & 20.726  &   0.0227 & 0.06777     \\
2454838.68496  & 20.562  &   0.0244 & -0.04762  \\
2454839.70064  & 20.936  &   0.0185 & -0.08634  \\
2454890.61491  & 20.637  &   0.0187 & 0.00276    \\
2454894.68743  & 20.518  &   0.0290 & -0.02356  \\
2454895.69238  & 20.894  &   0.0179 & 0.01631     \\
2454896.66180  & 21.041  &   0.0145 & -0.03853  \\
2454897.65715  & 20.674  &   0.0175 & -0.04530  \\
2454898.66015  & 20.544  &   0.0197 & -0.02811  \\
2454939.53373  & 20.578  &   0.0164 & -0.01384  \\
2454940.52747  & 20.642  &   0.0154 & 0.03494    \\
2454941.53928  & 20.996  &   0.0156 & 0.02525    \\
2454942.52421  & 20.899  &   0.0169 & -0.08243  \\
2454943.53594  & 20.559  &   0.0191 & 0.05531     \\
2454944.52955  & 20.798  &   0.0155 & -0.02327  \\
\hline\\
\end{tabular}
\end{center}
\label{rvtable}
\end{table}

Thirty-four radial velocity measurements were obtained with the CORALIE spectrograph on the 1.2m
Euler telescope  \citep{baranne,wasp4}.  The stable, temperature controlled, high-resolution echelle 
spectrograph has a resolution of $R\sim 55000$ over the spectral region from $3800 - 6800$~\AA.  
The WASP-19 spectra, obtained between 29 May 2008 and 23 April 2009, were processed through 
a slightly updated version of the CORALIE data reduction pipeline.  In addition to the
standard pipeline described in \citet{baranne}, we corrected for the blaze 
function and scaled the fitted cross-correlation region to match the full-width half maximum
of the object.  The final radial velocity (RV) values were obtained by cross-correlating
the spectra with a G2 template mask.  Table~\ref{tab:rvdata} presents the radial velocity 
measurements of WASP-19 at each Barycentric Julian date, the 1$\sigma$ Poisson errors 
on the velocities, and the line bisector span measurements \citep{gray,bisec2}.  
Based on our experience, we adopt uncertainties on the line bisector measurements of 
twice the measured RV errors.

\begin{figure}
 \centering
 \includegraphics[angle=0,width=\columnwidth]{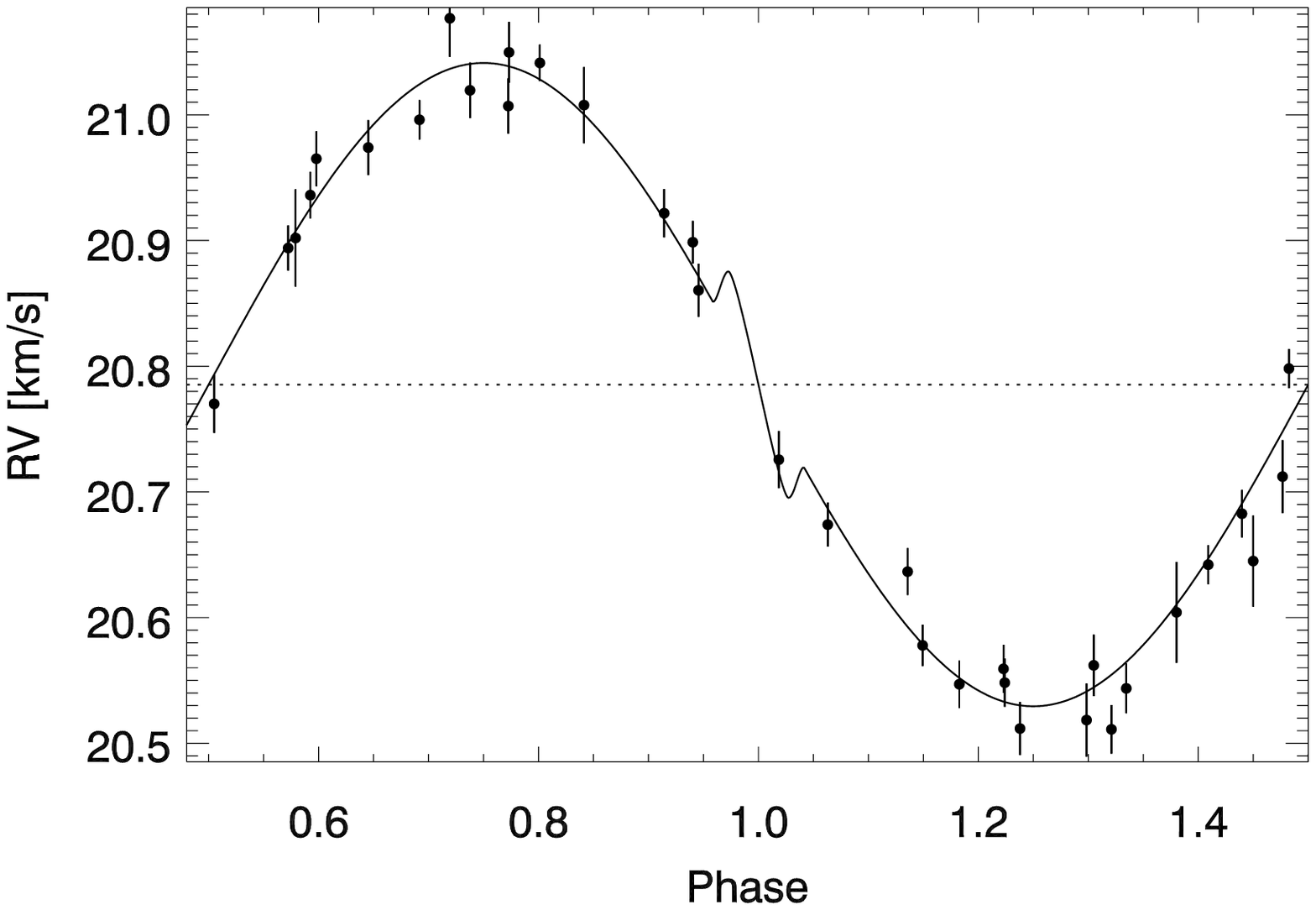}
 \caption{Radial velocity curve of WASP-19 phase-folded with
 the ephemeris given in Table~\ref{tab:params}.  Overplotted is the best fitting model
 curve obtained from a combined analysis of the photometric and spectroscopic data.
 }
 \label{fig:rvcurve}
\end{figure}

\begin{figure}
 \centering
 \includegraphics[angle=90,width=\columnwidth]{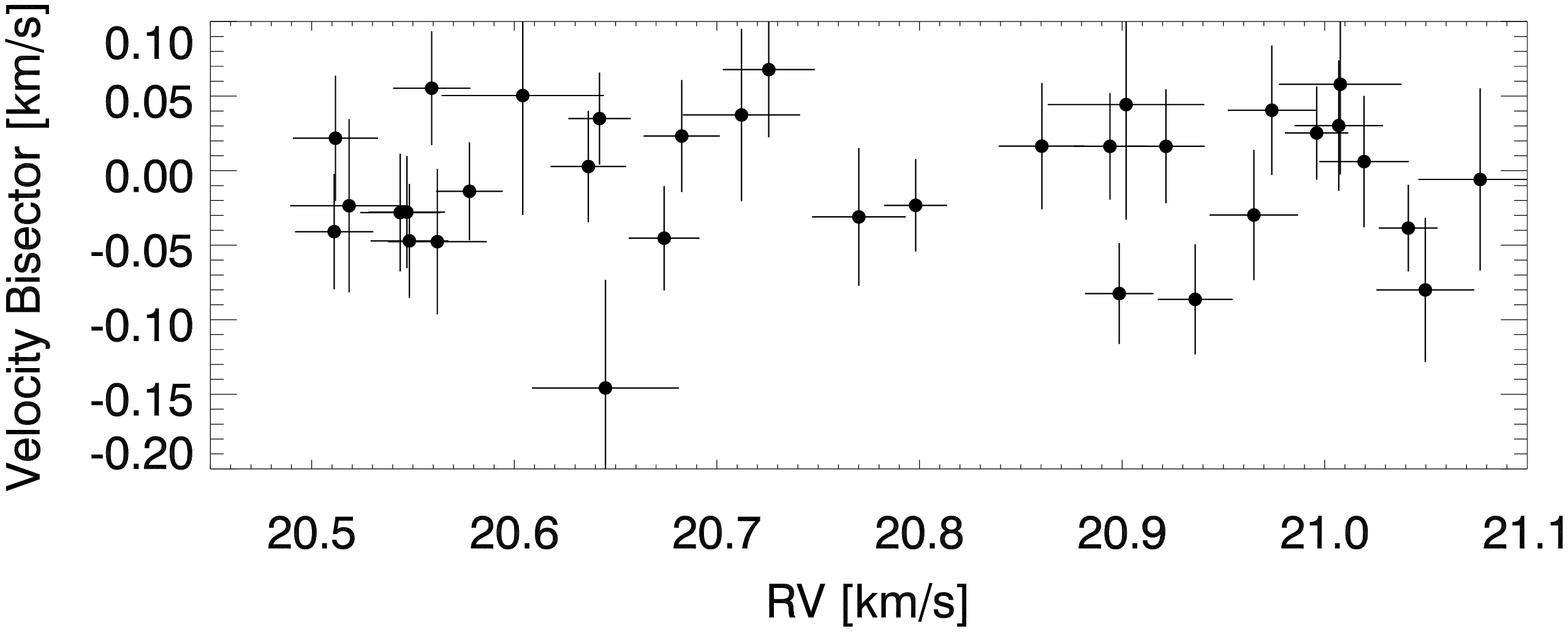}
 \caption{Line bisector span versus radial velocity.  
  The uncertainties on the line bisector values are double the values on the 
  radial velocity measurement.  There is no correlation between the line 
  bisector span measurements and radial velocity which rules out 
  star spot variations or a blended eclipsing binary as the cause for the velocity variations.  }
 \label{fig:bisec}
\end{figure}

The radial velocity of the star varies sinusoidally 
with the same period measured from the photometry (see Figure~\ref{fig:rvcurve}).  
In addition, the line bisector spans, which are used to discriminate
spot induced velocity variations and the effects of line-of-sight 
binarity, show no correlation with radial velocity within the uncertainties.  
The slope of the bisector versus RV (Figure~\ref{fig:bisec}) 
is  $-0.006\pm 0.037$, and the bisector measurements 
have an r.m.s.\ scatter of $\sim 50$~m~s$^{-1}$.  

Although all existing photometric and spectroscopic data suggests WASP-19 
is orbited by a short period transiting extra-solar planet, we explore the possibility
of a false positive detection.  In general, it is difficult 
to mimic the photometric and spectroscopic observations of a transiting planet 
without showing a visible second star in the spectrum, a significant bisector trend, 
and/or inconsistencies between the transit duration and host star spectral type.  
Starspots can cause periodic low-amplitude RV variations \citep[e.g.][]{huelamo,desort}
but not photometric transits as well, thus a single star blended with a 
fainter stellar eclipsing binary (EB) is the preferred false positive scenario
for transiting planets.  Unfortunately, no comprehensive simulations have been 
performed which model the expected RV variations, eclipse shapes, and bisector 
slopes for different blended EB scenarios, and performing such simulations is 
beyond the scope of this discovery paper.  
Instead, we explore the blended EB scenario through qualitative reasoning.

In order to produce a flat-bottomed, 2\% transit, as seen for WASP-19, 
the flux ratio of the EB compared to WASP-19 would have to be small enough
that the EB was undetectable as a peak in the cross-correlation function. 
However, it could not be so small that the necessary unblended eclipse depth 
would require nearly equal sized EB components and therefore, a V-shaped eclipse 
(i.e.\ $0.05 < F_{EB}/F_{W19} < 0.2$).
Although the eclipsing star would have to be small (and presumably less massive) 
compared to its primary to create the flat-bottomed eclipse, the RV
amplitude of the visible EB primary would still be 40-90~km~s$^{-1}$ for
all reasonable mass ratios due to the high inclincation angle needed to eclipse,
the short orbital period of the observed transits, and the relatively deep transit
(i.e.\ brown dwarf mass eclipsing objects would produce much shallower transit depths).
Therefore, the RV variations of the EB could not be hidden within the WASP-19 spectral
features given the resolution of the CORALIE data.

Furthermore, a short period stellar EB would almost certainly be
tidally synchronized with \vsini~$\sim 40-80$~km~s$^{-1}$.  
In the analysis of HD~41004~A \citep{santos}, which exhibits planet-like RV variations
due to a blended M~dwarf + brown dwarf spectroscopic binary (SB), the bisector
correlation is most dependent on the width of the visible SB component.
This system, in which the SB is only 3\% as bright as the single star 
and the width of the SB cross correlation function is $\sim 8$~km~s$^{-1}$, shows
a significant bisector correlation (slope of 0.67).  According to their
simulations, higher rotational broadening would produce an even greater 
bisector slope.  Therefore, we would expect WASP-19 to show a significant
bisector trend if the RV signal was caused by a rapidly rotating, 
blended EB, rather than a transiting planet.  

Finally, we see no difference in the eclipse depth of the z-band FTS
transit compared the SuperWASP transit taken in a bluer, V+R filter.  
Although this is not a strong constraint given the scatter in the SuperWASP 
photometry, an eclipse by a stellar object could show eclipse depth
variations in different filters which we do not see.
In summary, we conclude that the existing photometric and radial velocity variations of 
WASP-19 are most likely due to the presence of a transiting extra-solar planet.  

%Furthermore, \citet{desort} perform initial simulations that examine 
%the effect of starspots on radial velocity variations and bisector span
%measurements in the context of extra-solar planet host stars.  

\section{Analysis}
\label{sec:analysis}

A combined analysis of the radial velocity curve and light curve of a transiting
planet host star will provide direct measurements of the mass and radius of its orbiting planet
with one additional constraint, e.g.\ the stellar mass. 
Below, we describe the determination of the stellar mass and other host star properties.

\subsection{Spectroscopic parameters}
\label{sec:specparms}

The individual CORALIE spectra were co-added into a single
higher signal-to-noise (S/N) spectrum which was then used to measure the stellar
temperature, gravity, metallicity, $\vsini$ and elemental abundance 
information through comparisons with the synthetic spectra of
\citet{castelli}.  The results of the analysis are listed in 
Table~\ref{tab:specparms}.  The quoted uncertainties directly
correlate with the modest signal-to-noise of the co-added spectrum (S/N~$\sim 70$).

The spectral synthesis technique is described 
in detail in \citet{wasp6}, therefore we only give a short 
summary here.  The H$\alpha$ line was the primary temperature determinant,
while the Na~{\sc i}~D and Mg~{\sc i}~b lines were used as surface
gravity (\logg) diagnostics.   By measuring the equivalent width of 
several clean and unblended metal lines, we derived 
abundances for the elements listed in Table~\ref{tab:specparms}.    
Due to the spread in abundance measurements, we adopt
an overall metallicity for the system of [M/H]$=0.1\pm0.1$.
The microturbulence ($1.1 \pm 0.2$~km~s$^{-1}$), which directly affects the abundance measurements,
was derived from the Fe~{\sc i} lines using Magain's (1984) method.  

The projected rotational velocity ($\vsini$) was determined by fitting the
profiles of several unblended Fe~{\sc i} lines.  We accounted for line broadening
due to the instrumental FWHM (0.11\AA) which was determined from telluric
lines around 6300\AA, and the macroturbulence (2~km~s$^{-1}$) which was based on the 
tabulation by \citet{grayb}.  Finally, we do not detect Li~{\sc i} in the stacked spectrum, 
and can therefore only put an upper limit on the abundance of this element of
$\log A({\rm Li}) < 1 $.

\begin{table}
\caption{Stellar properties of WASP-19 obtained from the NOMAD catalogue and 
derived from our analysis of the spectra and light curves.}
\begin{center}
\begin{tabular}{cc}
Parameter    & WASP-19 \\
\hline\\
${\rm RA (J2000)}$    &  09:53:40.08         \\
${\rm Dec (J2000)}$   & -45:39:33.0         \\
${\rm J}$     &  $10.911\pm 0.026$   \\
${\rm H}$     &  $10.602\pm 0.022$   \\
${\rm K}$     &  $10.481\pm 0.023$   \\
${\mu_{\rm RA}}$ & $-41.3\pm2.5$ mas yr$^{-1}$  \\
${\mu_{\rm DEC}}$ & $16.5\pm1.9$ mas yr$^{-1}$ \\
${\rm U}$     &  $-49^{+21}_{-15}$ km s$^{-1}$  \\
${\rm V}$     &  $-25^{+2}_{-2}$   km s$^{-1}$ \\
${\rm W}$     &  $-13^{+7}_{-10}$  km s$^{-1}$ \\
              &                      \\
\teff      & 5500 $\pm$ 100 K \\
\logg      & 4.5 $\pm$ 0.2 \\
\mictrb    & 1.1 $\pm$ 0.1 \kms \\
\vsini     & 4 $\pm$ 2 \kms \\
{[Fe/H]}   &   0.02 $\pm$ 0.09 \\
{[Si/H]}   &   0.15 $\pm$ 0.07 \\
{[Ca/H]}   &   0.12 $\pm$ 0.15 \\
{[Ti/H]}   &   0.13 $\pm$ 0.12 \\
{[Ni/H]}   &   0.10 $\pm$ 0.08 \\
log A(Li)  & $<$1. \\
           &   \\
$\rho_*$ & $ 1.13\pm 0.12$ $\rho_{\odot}$ \\
$P_{rot}$ & $ 10.5\pm 0.2 $~days
\\
\hline\\
\end{tabular}
\end{center}
\label{tab:specparms}
\end{table}
 
\subsection{Host star mass}

By comparing the effective temperature, metallicity, and mean stellar density ($\rho_*$) of WASP-19 to
theoretical stellar models, we determine its mass.
The stellar density is dependent on the shape of the transit and largely independent of any
assumptions or models \citep{seagermaellen}.  Thus, after deriving the spectroscopic parameters, we model the WASP-South
and FTS transit light curves of WASP-19 using the Markov-chain Monte Carlo (MCMC)
routined described in \citet{mcmc} to derive the stellar density and its uncertainty.  
The code uses the MCMC approach
to simultaneously solve for the orbital and physical properties of the star-planet system. 
We apply the limb darkening coefficients of \citet{claret00,claret04} for the appropriate temperature
of the star and wavelength of the light curves.   We note that the eccentricity value
has an effect on the stellar density determination, thus we first
allow the eccentricity to be a free parameter.  Since the resulting eccentricity value gives
a non-signficant $1.5\sigma$ detection, we solve for the stellar density a second time
while fixing the eccentricity to be zero. 

\begin{figure}
 \centering
 \includegraphics[angle=0,width=\columnwidth]{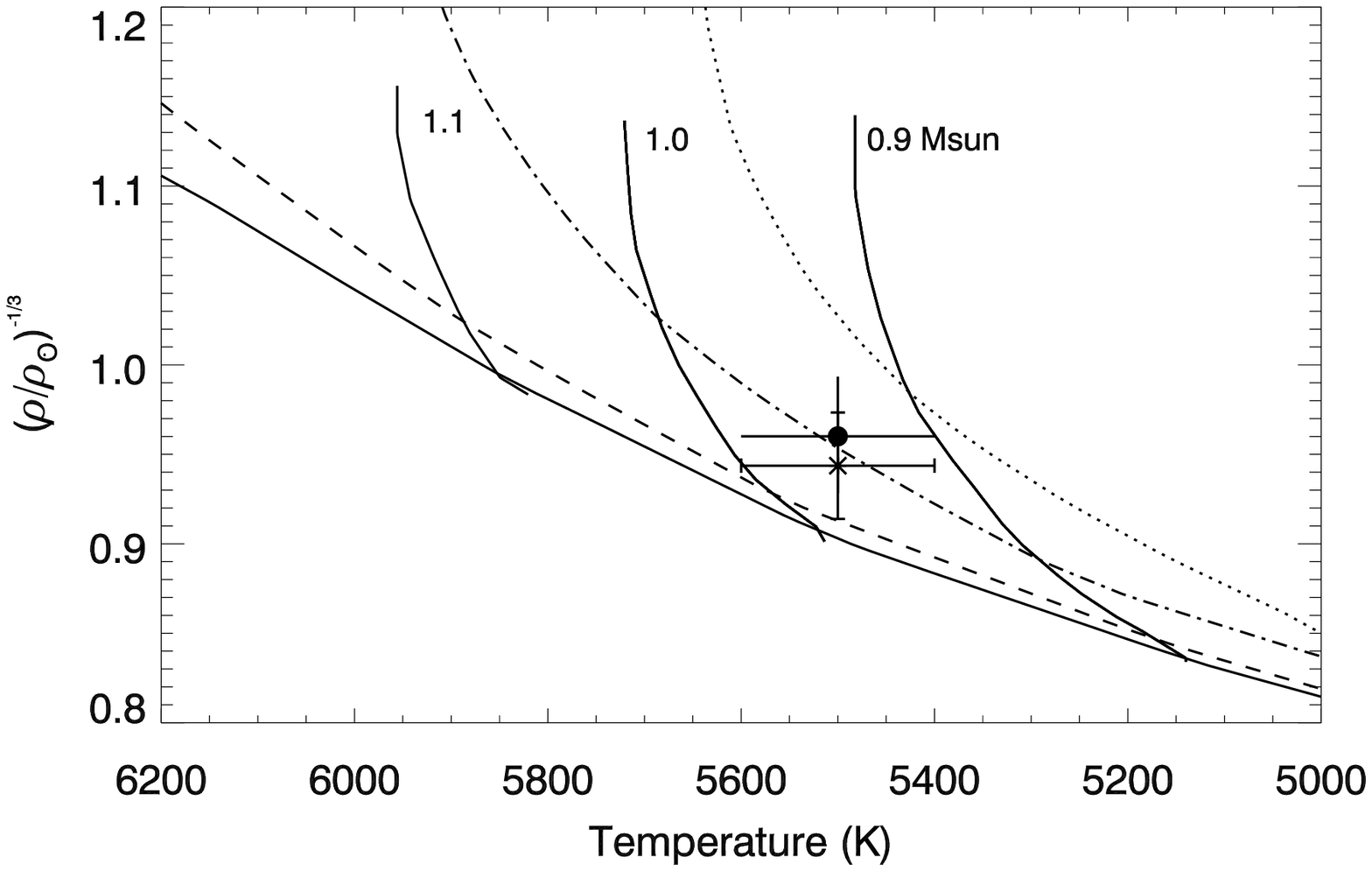}
 \caption{Modifed Hertzprung-Russell diagram comparing the stellar density and temperature of 
  WASP-19 to theoretical stellar evolution tracks by \citet{girardi00} interpolated at a 
  metallicity of [M/H]=+0.1.  The solid circle shows the result when the eccentricity is zero, and
  the asteriks shows the value for the stellar density if the eccentricity is a free floating parameter.  
  The mass tracks are labelled and the isochrones are 0.1 (solid), 1 (dashed), 5 (dot-dashed), and 10 (dotted) Gyr. 
  }
 \label{fig:evol}
\end{figure}

Interpolating among the \citet{girardi00} stellar evolution tracks as described in \citet{wasp12}, we derive
a mass for the host star of $M_*=0.95^{+0.09}_{-0.10} M_{\odot}$ using the circular orbit solution.
The non-circular orbit value for the stellar density results in a mass of 0.96~$M_{\odot}$, well within
the existing error bars (adopted from the non-circular solution).  
Figure~\ref{fig:evol} shows a plot of the position of
WASP-19 in a modified Hertzprung-Russell diagram as compared to the theoretical tracks.  
The errors on the stellar mass are dominated by the uncertainty on the metallicity.  
Here, we take into account uncertainties on the stellar temperature, metallicity and density. 
We do not include systematic uncertainties on the chosen stellar model  
which are difficult to determine, but we might expect this additional 
uncertainty to be at least $\sim 4$~\%.
\citet{southworth} finds that with stellar parameters measured at the limit of our technological and theoretical
ability for HD~209458b, the variation in the mass determination using four different stellar models is $\sim 4$\%.

\subsection{Host star age}

The isochrone fitting also allows for deriving an age estimate for WASP-19.
According to these stellar models, WASP-19 is a main-sequence star 
with an age of $5.5^{+9.0}_{-4.5}$~Gyr when we adopt the zero eccentricity value for
the stellar density.  This essentially places a weak constraint on the stellar age to be
$\gtrsim 1$~Gyr.  If we adopt the value for the stellar 
density from the solution when the eccentricity is a free parameter
(non-significant $1.5\sigma$ result), we are unable to place any constraints on the 
stellar age from the isochrones.  The depletion of lithium in the atmosphere of a
G8V star can also be used as an age indicator.
However, the non-detection of Li~{\sc i} in WASP-19 again gives only a weak constraint
on the age suggesting the star is older than the Hyades \citep[0.6~Myr,][]{sesitorandich}.

We, therefore, examine the three dimensional velocity of WASP-19 as compared to theoretical
Galactic model stars to estimate the probability that WASP-19 is part of a young disk population.  
Using the catalogue proper motions and measured systemic radial velocity,  we calculate its
$U$, $V$, and $W$ space motion (given in Table~\ref{tab:specparms}) compared to the Sun.
We estimate the distance of WASP-19 
to be $250^{+80}_{-60}$~pc using the measured spectral type, G8V, with absolute magnitude, 
$M_v=5.6$ from \citet{gray} and V magnitude from the NOMAD catalogue ($V=12.59$).
We adopt a $\pm 0.2$ uncertainty on the magnitude measurement as it is not given in the catalogue
and a generous error of $\pm 2$~subclasses in spectral type to determine the uncertainty 
on the distance estimate.  We select a set of model main sequence stars of spectral types F7-K7 
in a small volume around WASP-19 ($ l = 273\pm 3^{\circ}$, $ b = 7\pm 3^{\circ}$, distance=150-400~pc) 
using the Besan\c{c}on Galactic model \citep{besancon}.  We generate 100 resolutions of 
the simulation which provide the heliocentric velocities, metallicities, and population classes
for over 500,000 model stars.  Of the model stars with metallicities of 0.0-0.2 dex and 
with space motions within the errors of the calculated $U$,$V$, and $W$ values of WASP-19,
35\% of the model stars have ages of $< 1 $~Gyr (population class 1 and 2), 59\%
have ages of 1-5~Gyr (population class 3-5) and 6\% have ages of 5-10~Gyr.
Contributions from the thick disk, halo and bulge populations are negligible.  This analysis 
suggests WASP-19 has a 65\% probability of being older than 1~Gyr.

In summary, we present three different age dating techniques which all suggest
WASP-19 is older than $\sim 1$~Gyr, however a precise
age for the star cannot be determined from the existing data.  

\subsection{Host star rotation period}

%We also investigate the stellar rotation which is a possible age indicator
%for a main sequence star of this spectral type \citep{barnes}.  

We search for variability in the WASP-South light curves of WASP-19 caused by asymmetric starspots on the 
photosphere which modulate the flux.  Any starspots will rotate in and out of view 
with the stellar surface, such that the period of the variability gives
the rotation period of the star.  To exhibit rotational variability, 
the star must have a sufficient coverage of starspots to create a variable 
brightness signal which is detectable given the scatter in the photometric
data.  Furthermore, starspots evolve on timescales of weeks or months causing
the amplitude and phase of the variability to change, therefore,
each season of WASP-South data was examined independently.

\begin{figure}
 \centering
 \includegraphics[angle=0,width=\columnwidth]{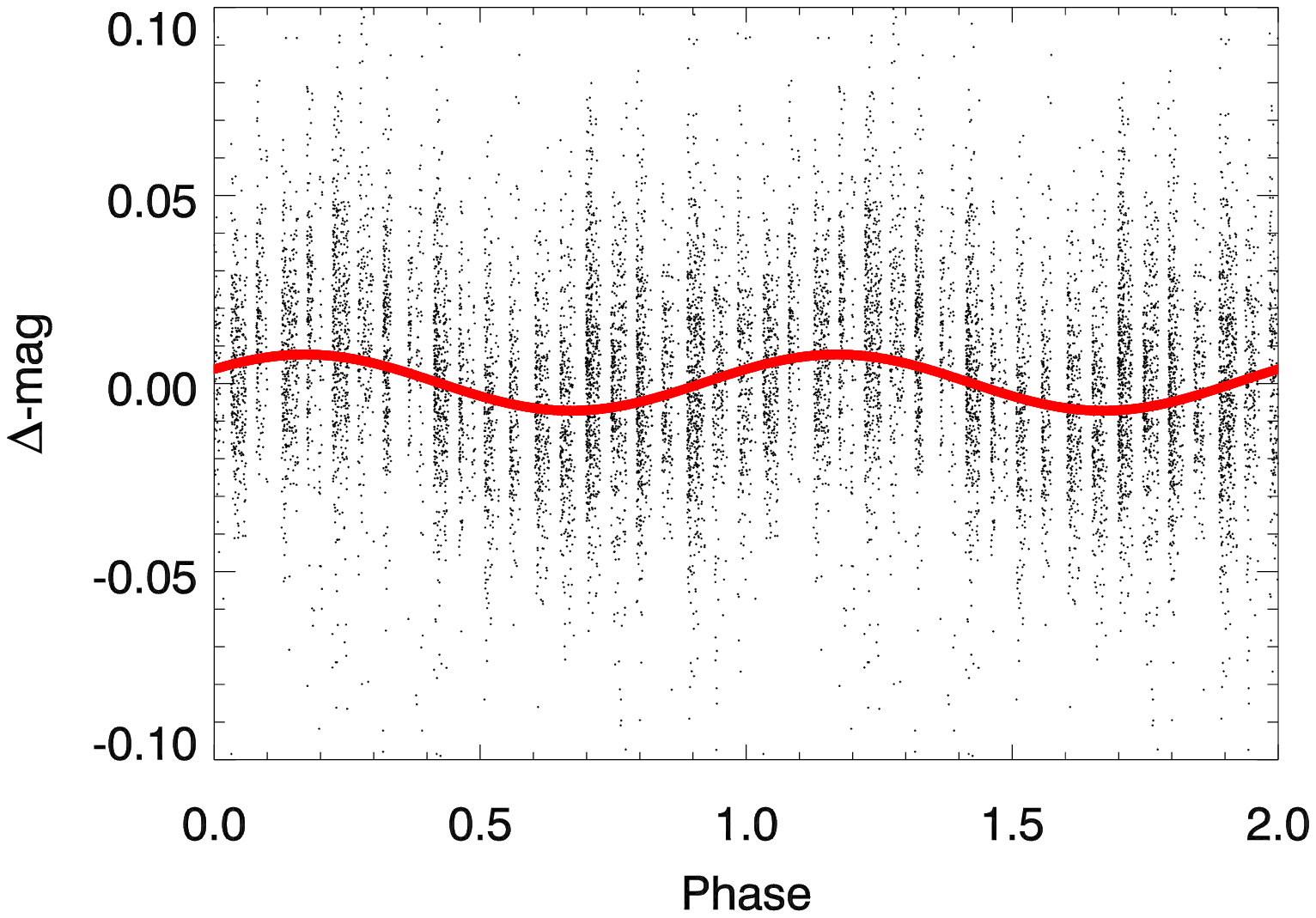}
 \caption{WASP-South light curve data from 2007 phase-folded on the rotation
  period detected in the sine fitting, $P_{rot}=10.5$~days.}
 \label{fig:rotvar}
\end{figure}

\begin{figure}
 \centering
 \includegraphics[angle=0,width=\columnwidth]{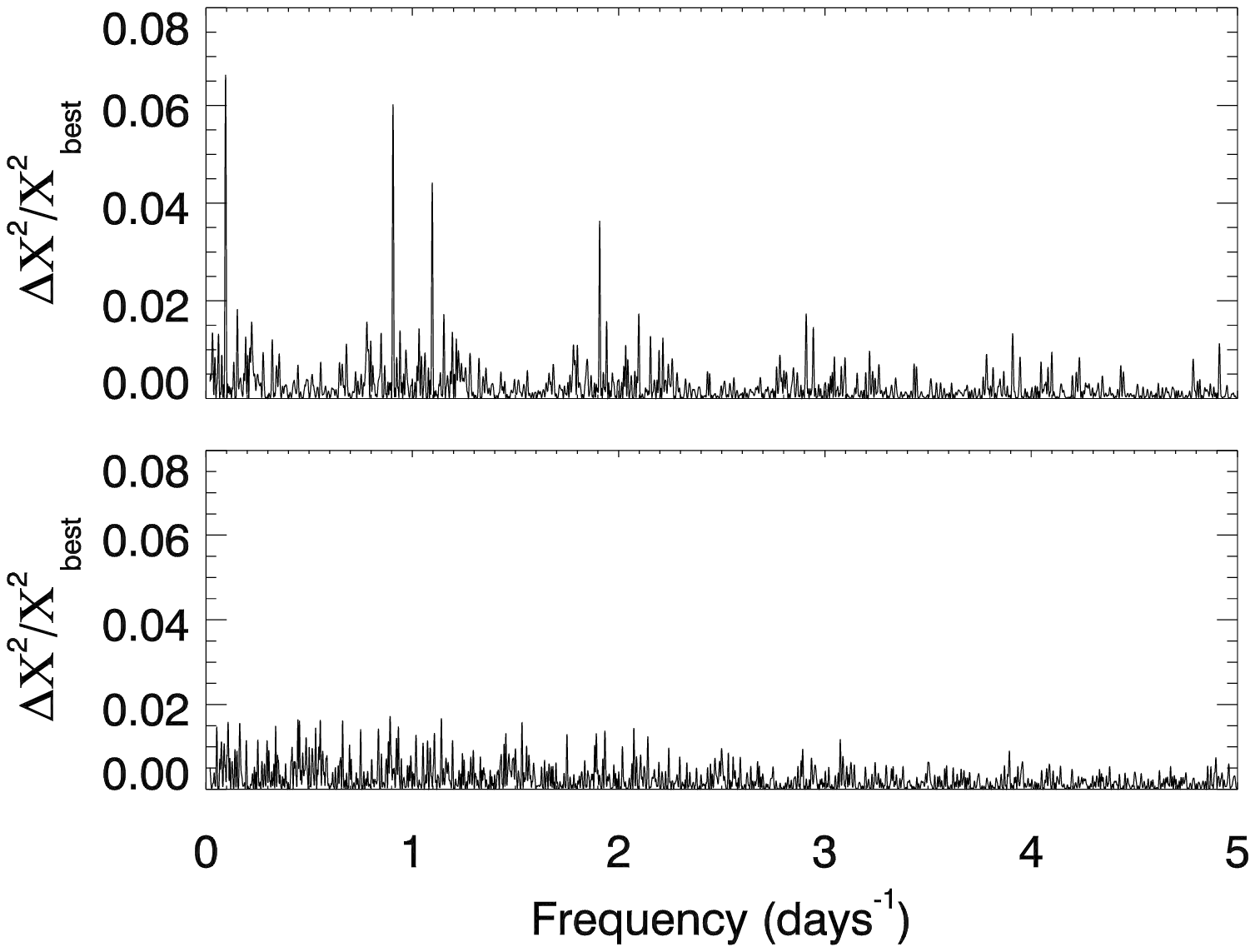}
 \caption{Top: Periodogram, $\Delta \chi^2/\chi^2_{best}$ versus frequency, resulting from fitting
  a sine wave to the 2007 WASP-South light curve.  The peak in the
  periodogram has a $FAP=1\times10^{-7}$ and occurs 
  at $P_{rot}=10.5$~days indicating the rotation period of the star.
  Bottom: Periodogram resulting from fitting a sine wave to the shuffled
  2007 light curve in which the periodic signal was destroyed, but the red \citep{Pont06} and
  white noise were preserved. We adopted a $FAP=0.95$ for the highest peak. }
 \label{fig:pdgram}
\end{figure}

To detect the rotational variability, we determine the improvement in $\chi^2$ over
a flat, non-variable model when a sine wave of the form  
$y = a_0 + a_1 sin(\omega t + a_3)$ is fit to each season of the 
WASP-South data phase-folded at a set of trial periods, $P_{rot}=2\pi/\omega$.  
We subtract all transits from the light curves using the model derived 
from the parameters in Table~\ref{tab:params} before fitting the sine curve model.
We test periods between $0.2-50$~days and
find a strong periodic signal in the 2007 data with $P_{rot}=10.5$~days
and amplitude, $a_1=$~7.6~mmag.  The phase-folded light curve and periodogram
of normalized $\Delta\chi^2$ values are shown in Figure~\ref{fig:rotvar} and Figure~\ref{fig:pdgram} (top),
respectively.  In the 2008 data, we also detect a weaker signal 
with a similar period, $P_{rot}=10.6$~days, and with an amplitude of 3.6~mmag. 

To assess the veracity of the sinusoidal signal detected in the 2007 data, we determine
the significance and the false alarm probability (FAP) following \citet{zechkurst}
(employing the residual variance normalization).
The equations include, $N$, the number of
independent data points, and $M$, the number of independent frequencies, as well as the
measured peak value in the periodogram.  According to
\citet{cumming}, the number of independent frequencies can be approximated by the duration
of the time-series data times the difference between the highest and lowest frequencies
tested (here $M=753$).  The number of independent data points depends on the level of red noise in the 
light curve.  We empirically determine $N$ by generating a new light curve in which
the red noise is preserved, but the periodic signal is destroyed.  We randomly
re-order the individual nights of data, so that the integer part of the 
light curve time values are shuffled, but the fractional parts (containing the red noise) remain the same.
We then run the sine fitting program on the 
shuffled light curve and find $N=664$ by assuming the highest peak in the resulting
periodogram (Figure~\ref{fig:pdgram} (bottom)) has a 95\% probability of being false ($FAP=0.95$).  
Using our calculated $N$ and $M$, we then apply the equations in \citet{zechkurst} to the
results of the sine fitting on the original 2007 light curve 
and find a highly significant 
periodic signal with a $\rm{Prob}(p>p_{best})= 1.4\times10^{-10}$ and $FAP=1\times 10^{-7}$.
Thus, we adopt a rotation period for WASP-19 of $P_{rot}=10.5$~days.  

We calculate the error on this measurement using the formula
in \citet{horne} and find $\sigma_P=0.08$, but given the variation in period
values we measure using two other techniques (Lomb-Scargle and auto-correlation), we 
find this to be underestimated and suggest  $\sigma_P=0.2$~days is a more realistic 
uncertainty.  Finally, we note that the 10.5 day rotation period of the star measured via the photometry 
is consistent with the less precise $\vsini$
value of $4 \pm 2$ \kms which corresponds to a rotational period of between 8-24~days
(assuming the spin axis of the star is perfectly aligned with the orbital axis).

%As there are no obvious systematics
%in the photometry which would cause a spurious variability signal with a period 
%of $\sim 10.5$~days in two independent seasons of SuperWASP data, 
%and the periodic signal in the 2007 data is highly signfant
%we consider the detection to be valid and adopt a rotation period
%for the star of P=10.5~days.  
%Finally, we use two additional techniques to look for a periodic signal
%in the 2007 data.  We use a standard Lomb-Scargle periodogram and find a
%signficant peak at $P_{rot}=10.7$~days.  We also employ an auto-correlation
%technique \citep{acc_coma} which determines the best period to be $P_{rot}=10.4$~days.

The rotation period of a main sequence star can also be used to estimate its age.
According to the \citet{barnes} gyrochronology relationship, a rotation period of $P_{rot}=10.5$~days 
corresponds to an age of 500-600 Myr for a G8V star like WASP-19 with B-V=0.74.  
This value is not consistent with the older age inferred from the isochrones, lithium, and 
space motion.  In \S~\ref{sec:discuss}, we discuss the implications of the possible 
discrepancy between the different age indicators.
%Therefore either WASP-19 is actually younger than 1~Gyr or some physical
%phenomena has caused WASP-19 to rotate faster than is expected given its age.
%We discuss the causes and implications for the spin up of the star in \S~\ref{sec:discuss}.

\subsection{Planet properties}

We solved for the properties of the star-planet system by running
an MCMC code \citep{mcmc} using as inputs the WASP-South light curve, the FTS
light curve, the CORALIE radial velocity curve, and the stellar mass. 
Initially, we allowed the $e$cos$\omega$ and $e$sin$\omega$ to be free parameters, 
but the low, non-zero values that resulted were not significant.  Therefore, we also
solved for the parameters of the system while forcing the planet to be on 
a circular orbit.   The difference in stellar
and planet properties derived for the circular and non-circular orbit cases is 
negligible and within the $1\sigma$ uncertainties on all parameters.
However, for completeness, we provide the resulting solutions for both cases.
Table~\ref{tab:params} gives
the final model parameters for the transit and radial velocity curves as well as
the physical properties of the star-planet system.  

Note, we do not remove the rotational
variability from the light curves before deriving the final system  parameters.  
The model fit is dominated by the FTS data, and the low-amplitude 10.5~day rotational variability
is negligible on the 3.3~hour timescale of this light curve.  We confirm this by fixing
the orbital period to the value found when analysing all the light curve data and fitting
just the FTS light curve and radial velocity data.  All the resulting light curve parameters 
and physical system parameters are well within their reported 1$\sigma$ uncertainties.  
Therefore, we do not `pre-whiten' the WASP-South data before performing the final 
transit model fits on all the existing photometric data.  However, it is important to note
that we cannot determine if any starspots were present on WASP-19 during the FTS
observations which might affect the resulting parameter determinations because of the 
relatively small amount of out-of-transit data.  Additional high quality transit 
data would allow for investigating any variations in the derived parameters due to
starspots.

Finally, to derive the final parameters, we use the 4-coefficient limb darkening model by 
\citet{claret04,claret00}
with the ATLAS model atmospheres, an effective temperature of 5500 K, and a \logg of 4.5.
However, we tested a range of limb darkening coefficients with different temperatures,
\logg\ values, and model atmospheres and found the resulting parameters to be highly robust 
to changes in these coefficients.  We suspect this is due to the fact
the FTS z-band data dominates our parameter results, and this filter
is not as susceptible to limb darkening uncertainties as bluer filter data.

\begin{table*}
\begin{center}
\caption[]{WASP-19 system parameters and 1$\sigma$ error limits derived
from the MCMC analysis.}
\label{tab:params}
\begin{tabular}{lcccl}
\hline\\
Parameter & Symbol & Value ($e$ fixed) & Value ($e$ free) &  Units \\
\hline\\
Transit epoch (BJD)     & $ T_0  $ & $ 2454775.3372^{+ 0.0001 }_{- 0.0002 } $ & $2454775.3372^{+ 0.0002}_{- 0.0002}$  & days \\
Orbital period          & $ P  $ & $ 0.7888399^{+ 0.0000008 }_{- 0.0000008 } $   & $0.7888399^{+ 0.0000008 }_{- 0.0000008 }$  & days \\
Planet/star area ratio  & $ (R_p/R_s)^2 $ & $ 0.0203^{+ 0.0004 }_{- 0.0004 } $   & $0.0203^{+ 0.0004 }_{- 0.0004 }$ &  \\
Transit duration        & $ t_T $ & $ 0.0642^{+ 0.0006 }_{- 0.0006 } $           & $0.0643^{+ 0.0006 }_{- 0.0007 }$  & days \\
Impact parameter        & $ b $ & $ 0.62^{+ 0.03 }_{- 0.03 } $                   & $0.62^{+ 0.03 }_{- 0.03 }$  & $R_*$ \\
  &    &      &   &  \\
Stellar reflex velocity & $ K_1 $ & $ 0.256^{+ 0.005 }_{- 0.005 } $              & $0.256^{+ 0.005 }_{- 0.005 }$  & km s$^{-1}$ \\
Centre-of-mass velocity & $ \gamma $ & $ 20.78534^{+ 0.0002 }_{- 0.0002 } $      & $20.78535^{+ 0.0003 }_{- 0.0003 }$  & km s$^{-1}$ \\
Orbital semimajor axis  & $ a $ & $ 0.0165^{+ 0.0005 }_{- 0.0006 } $             & $0.0164^{+ 0.0005 }_{- 0.0006 }$  & AU \\
Orbital inclination     & $ I $ & $ 80.5^{+ 0.7 }_{- 0.7 } $                     & $80.8^{+ 0.8 }_{- 0.8 }$  & degrees \\
Orbital eccentricity    & $ e $ & $ 0 $ (fixed)                                  & $0.02^{+ 0.02 }_{- 0.01 }$  &  \\
Longitude of periastron & $ \omega $ & $ 0 $ (fixed)                             & $-76^{+ 112}_{- 23} $ & deg  \\
eccentricity $\times$ cos($\omega$) & $e$cos$\omega$ & $ 0 $ (fixed)                  & $0.004^{+ 0.009}_{- 0.009}$  &  \\
eccentricity $\times$ sin($\omega$) & $e$sin$\omega$ & $ 0 $ (fixed)                   & $-0.02^{+ 0.02}_{- 0.02} $ &   \\
  &    &      &   &  \\
Stellar mass & $ M_* $      & $ 0.96^{+ 0.09 }_{- 0.10 } $                       & $0.95^{+ 0.10 }_{- 0.10 } $ & $M_\odot$ \\
Stellar radius & $ R_*$     & $ 0.94^{+ 0.04 }_{- 0.04 } $                       & $ 0.93^{+ 0.05 }_{- 0.04 } $ & $R_\odot$ \\
Stellar surface gravity     & $ \log g_* $ & $ 4.47^{+ 0.03 }_{- 0.03 } $        & $ 4.48^{+ 0.03 }_{- 0.03 } $ & [cgs] \\
Stellar density & $ \rho_*$ & $ 1.13^{+ 0.09 }_{- 0.09 } $                       & $ 1.19^{+ 0.12 }_{- 0.11 } $ & $\rho_\odot$ \\
  &    &      &  &  \\
Planet radius & $ R_p $     & $ 1.31^{+ 0.06 }_{- 0.06 } $                       & $ 1.28^{+ 0.07 }_{- 0.07 } $  & $R_J$ \\
Planet mass & $ M_p $       & $ 1.15^{+ 0.08 }_{- 0.08 } $                       & $ 1.14^{+ 0.07 }_{- 0.07 } $ & $M_J$ \\
Planetary surface gravity   & $ \log g_p $ & $ 3.19^{+ 0.03 }_{- 0.03 } $        & $ 3.20^{+ 0.03 }_{- 0.03 } $ & [cgs] \\
Planet density & $ \rho_p $ & $ 0.51^{+ 0.06 }_{- 0.05 } $                       & $ 0.54^{+ 0.07 }_{- 0.06 } $ & $\rho_J$ \\
Planet temperature ($A=0$,F=1) & $ T_{\mbox{eq}} $ & $ 2009 ^{+ 26 }_{- 26 } $   & $ 1993 ^{+ 32 }_{- 33 } $ & K \\
\hline\\
\end{tabular}
\end{center}
\end{table*}

\subsection{Transit Timing}

We measured the heliocentric Juilian date of the mid-transit times for WASP-19b 
using the technique described in \citet{wasp17}
to search for variations which could indicate the presence of a outer planet \citep[e.g.][]{agol}.
In general, the SuperWASP transits are not precise enough to provide a useful 
constraint on the existence of a third body in the system, but
these data do rule out transit timing variations larger than $\sim 15$~minutes.

\section{Discussion}
\label{sec:discuss}

The most striking aspect of WASP-19b is its extremely short orbital period.  
With a period, $P=0.7888399 \pm 0.0000008$~days, WASP19b is the shortest period planet yet discovered.  
What physical processes in the evolution of the planet have lead to such a close separation? 
Did the initial migration process leave the planet in its extremely short period orbit 
or has there been subsequent evolution of the orbital separation?  
The observations presented here suggest the possibility that WASP-19b has been spiralling into 
its host star throughout its lifetime and has spun up its host star in the process.

Most transiting extrasolar planets will ultimately spiral into their host stars 
because there is insufficient total angular momentum in the systems to reach
a state of tidal equilibrium \citep{hut,rasio,jack09}.  However, the timescale for this evolution
is not well known because the stellar tidal quality factor, $Q_{s}^{\prime}$, is uncertain to at 
least three orders of magnitude.  Values between $10^6 < Q_{s}^{\prime} < 10^9$ are
reasonable \citep[][and references therein]{jackson08}.  
Furthermore, the planetary tidal quality factor, $Q_{p}^{\prime}$ is equally uncertain, but
it affects the evolution of the stellar spin and planetary orbital separation 
to a much lesser degree.

For WASP-19b, the ratio of total to critical
angular momentum is well below the limiting value of one, thus
ensuring the planet will ultimately collide with its host star, but
the lifetime of this evolution
ranges from 4~Myr to 4~Gyr depending on the value used for $Q_{s}^{\prime}$.
However, unlike most other transiting planets, we have measured the rotation period of WASP-19
which contributes additional information that can be used to place
tighter constraints on the overall lifetime of the planet and on the tidal quality factor of the star.

%A typical single G-dwarf with a mass of $M \sim 0.95 M_{\odot}$ will lose angular momentum 
%slowly over its lifetime due to magnetic braking.  At an age of 5~Gyr, the 
%star should have a stellar spin period of $\sim 25$~days, similar to that of the Sun.
%However, the measured rotation period of WASP-19 is much shorter than 
%is expected given its isochrone age.  

\begin{figure}
 \centering
 \includegraphics[angle=0,width=\columnwidth]{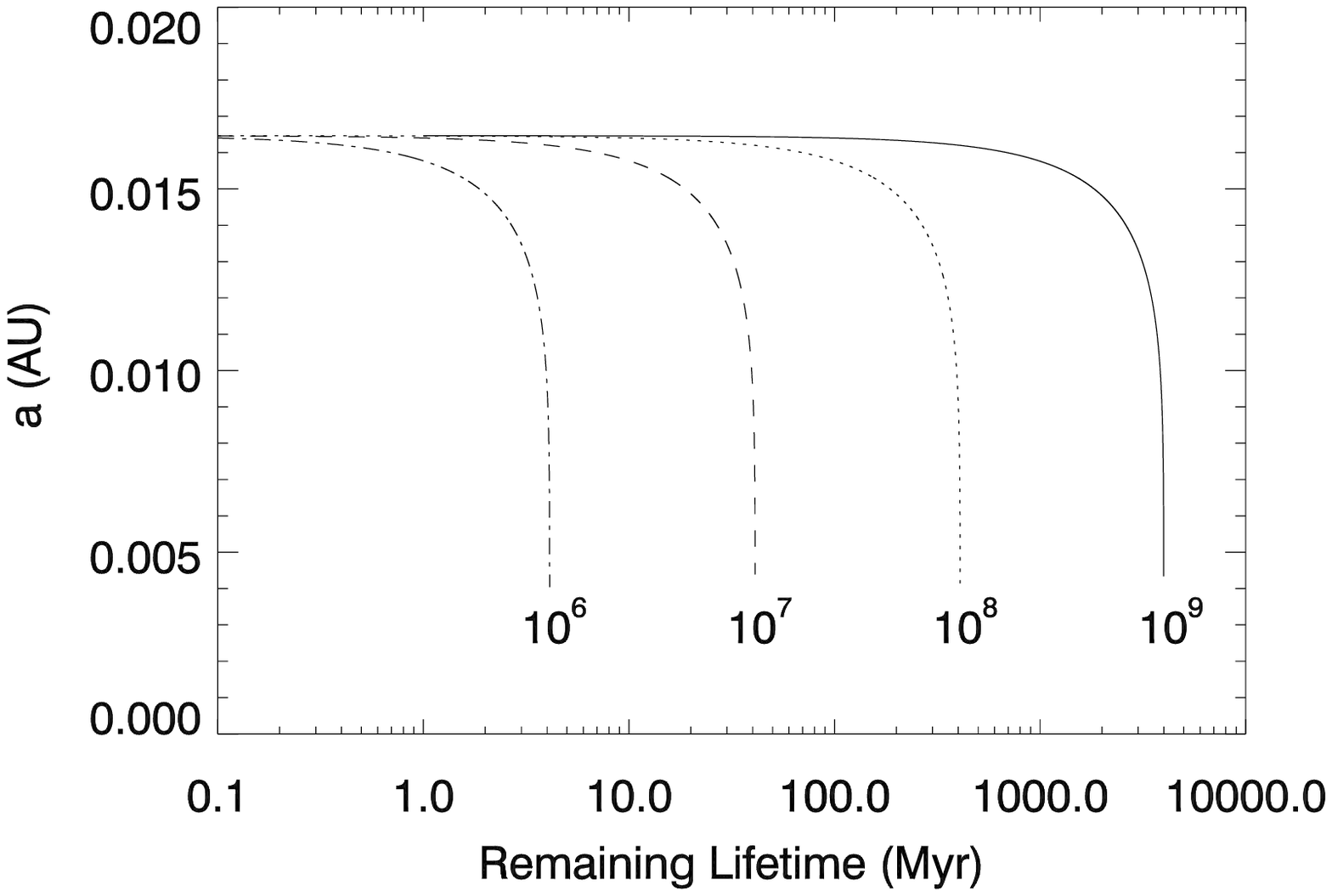}
 \caption{Orbital separation versus age of WASP-19b when evolved forward from the current
   state to the point when the planet overflows its Roche lobe.  Due to exchange of angular
   momentum through tides, the planet loses orbital angular momentum and spirals into the star.  
   The different lines which are labelled 
   correspond to different values for the stellar quality factor, $Q_{s}^{\prime}$.
   The remaining lifetime of the planet ranges from 4 Myr to 4 Gyr depending on the value
   of $Q_{s}^{\prime}$.}
 \label{fig:tides}
\end{figure}

\begin{figure}
 \centering
 \includegraphics[angle=0,width=\columnwidth]{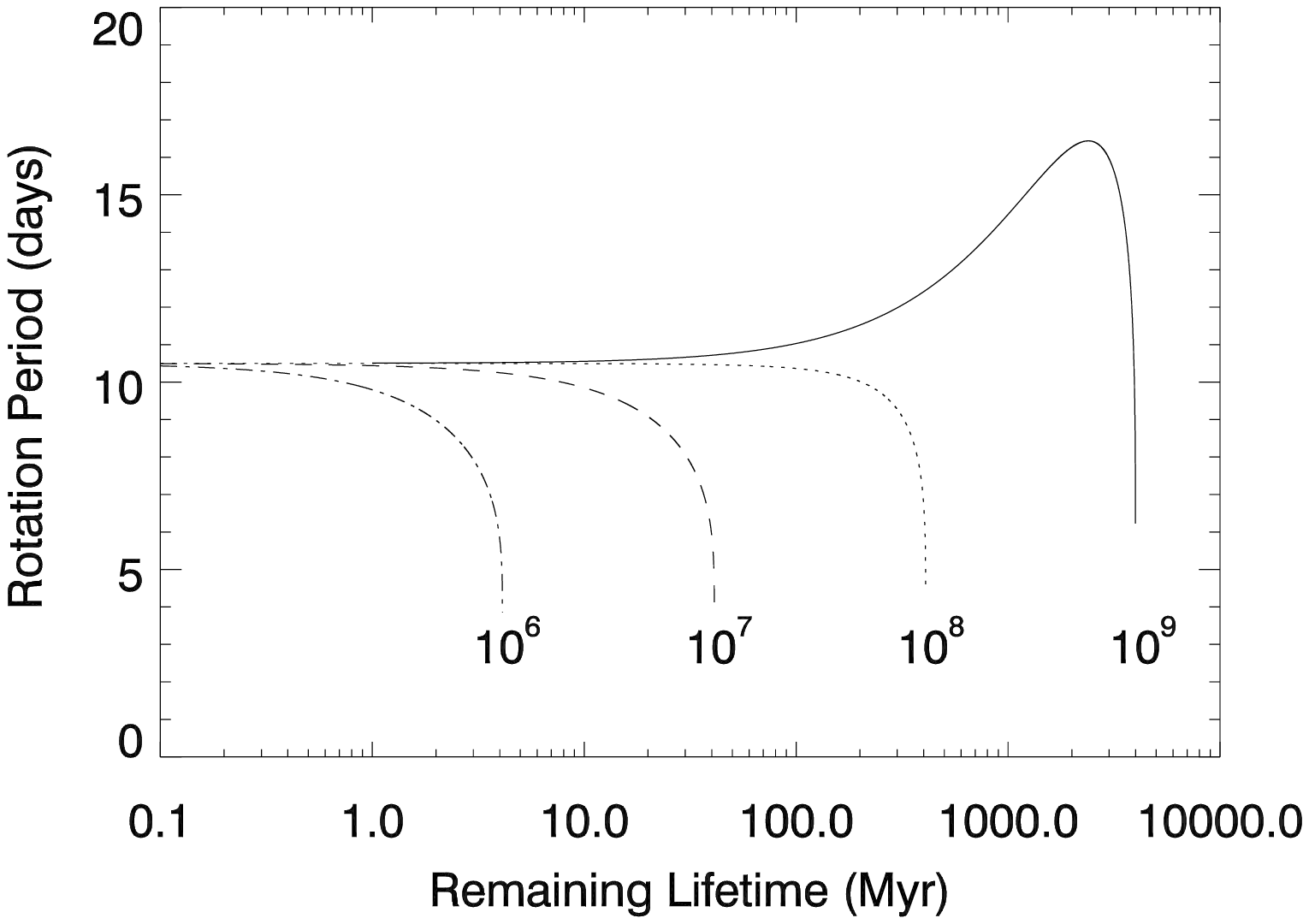}
 \caption{Stellar rotation versus age of WASP-19b when evolved forward accounting for
   tidal evolution.  The star gains rotational angular momentum and spins up as the planet
   spirals into the star, losing angular momentum.  The different lines are labelled for
   different values of $Q_{s}^{\prime}$.  For $Q_{s}^{\prime}=10^9$, the tidal evolution is
   slow enough that magnetic breaking due to stellar winds dominates the angular momentum
   of the star causing it to slow down prior to the eventual spiral in of the planet after
   $\sim 4$~Gyr.}
 \label{fig:spin}
\end{figure}

By integrating the coupled equations of orbital eccentricity and separation \citep{dobbsdixon}
while conserving the total angular momentum of the system,
we can model the future evolution of the planet as it spirals inward and further 
spins up the host star.  In the analysis,  we include magnetic braking which 
takes the form $\dot{\Omega} = -\kappa\Omega^3$
with $\kappa = 3.88 \times 10^{-7}$~s as evaluated using equation (12) of \citet{accandli}.  
%The other relevant stellar properties were interpolated for a star of $M = 0.95~M_{\odot}$ 
%from the values tabulated in the same paper.  
We use the current orbital separation derived from the MCMC analysis (Table~\ref{tab:params}) 
and the measured rotation period ($P_{rot}=10.5$~days) as initial conditions.
Figure~\ref{fig:tides} shows the future evolution of the planet in orbital separation,
and in Figure~\ref{fig:spin}, we plot the corresponding rotational evolution of the star.

The results show that in the $Q_{s}^{\prime} = 10^9$ case, the tidal interaction is so weak
that the star's future evolution is dominated by magnetic braking.  In this situation,
the relatively short rotation period could not have been caused by prior tidal interactions
with the planet and would instead suggest a young stellar age similar to that of the
Hyades ($\sim 600$~Myr).  Although it is possible WASP-19 is as young as the Hyades,
the isochrones analysis, the non-detection of lithium, and the stellar velocity all favor an older
age for the star.  Therefore, a lower value for $Q_{s}^{\prime}$ is preferred.

For all models with $Q_{s}^{\prime} < 10^8$, the star's spin period has already passed through
a maximum and is decreasing as the planet spirals in.   However, in the 
$Q_{s}^{\prime} = 10^6$ case, the remaining lifetime of the planet is extremely short (4~Myr).  
Therefore, the existing data appears to favor
values of $Q_{s}^{\prime}=10^7-10^8$ which are high enough to give significant life expectancy, but low
enough to have reversed magnetic braking and initiated spiral-in.
This scenario would allow for the extremely short period of the planet to be a consequence 
of further evolution in orbital separation after the initial migration process early in its
formation and evolution.  However, the $Q_{s}^{\prime}$ would have to be 1-2 orders
of magnitude greater than the nominal value of $10^6$ that is typically 
adopted \citep[e.g.][]{boden03,jackson08}.  

It is important to note that $Q_{s}^{\prime}$ is a simple parameterization of the 
complex physics involving the interaction between the tides raised by the planet on its star 
and the turbulent viscosity in the stellar convection zone and inertial wave modes 
excited in the stellar interior \citep{rasio,sasselov,ogilvielin}.  A more detailed
analysis of this system and others like it (e.g. WASP-18, OGLE-TR-56b, WASP-12)
will hopefully lead to a better understanding of the physics modulating
tidal dissipation in stars, since currently, there is no comprehensive theory
which is able to explain observations of both main sequence binary stars and close-in
extra-solar planets with regard to their tidal evolution \citep{rasio,sasselov,terquem,ogilvielin}.  

%Measurements of the 
%circularization timescale of binary systems \citep{latham,mm03} suggests a relatively
%strong tidal interaction and rapid circularization \citep{terquem}, whereas 
%a much weaker tidal dissipation is required for
%short period planets like 51~Peg~b \citet{rasio}, OGLE-TR-56b \citet{sasselov}, and WASP-19
%to even exist.  To fully understand and resolve these discrepances requires
%a more detailed analysis of this system and others like it (e.g. WASP-18, OGLE-TR-56b, WASP-12)

%Due to the extreme proximity to its host star, the planet's Roche radius is only $1.33\times$
%bigger than its radial size, thus it is on the verge of being tidally disrupted.

Finally, we note that the stellar flux incident on WASP-19b at the substellar
point is $3.64\times 10^{9}$~ergs~cm$^{-2}$~s$^{-1}$ placing it in the class of highly irradited
planets like OGLE-TR-56b \citep{ogle56,torres} and WASP-1b \citep{wasp1}.
Therefore, we expect the planet to have large secondary eclipse depths in the mid-IR, 
evidence of an atmospheric temperature inversion,  molecular emission features, and 
a large day/night contrast \citep{fortney08}.  The existence of a hot stratosphere can be 
tested using secondary eclipse measurements that are currently being 
obtained with Spitzer.  Eclipse measurements in the near-IR and optical z-band are also
possible given current technology \citep[e.g.][]{corot1}.  Furthermore, the density of WASP-19b 
is half that of Jupiter's ($\rho_p = 0.51 \rho_J$), so it is slightly bloated for its
mass, but not extremely so.  The high irradation, increased metallicity of the
host star, and dissipation of tidal energy are 
possible factors causing the enhanced radius \citep{boden03,fortney07,burrows07}.
 
In summary, WASP-19b is the shortest period transiting planet yet detected.
It has a mass, $M_{\rm pl}$=1.15~M$_J$ and radius, $R_{\rm pl}$=1.31~R$_J$.
The planet orbits a main sequence G-dwarf with a slightly super-solar metallicity and
a rotation period of $P_{rot}=10.5\pm 0.2$~days.  It is likely WASP-19b has been spiraling
into its host star over its lifetime and has spun up the star in the process.  A more precise age determination
of WASP-19 will allow us to confirm this and to place stronger constraints on the star's tidal
quality factor, $Q_{s}^{\prime}$.

\acknowledgements
The SuperWASP Consortium consists of astronomers primarily from the Queen's University Belfast,
St Andrews, Keele, Leicester, The Open University, Isaac Newton Group La Palma and
Instituto de  Astrof{\'i}sica de Canarias. The SuperWASP Cameras were constructed
and operated with funds made available from Consortium Universities and the UK's Science and
Technology Facilities Council.

\label{lastpage}

\end{document}